\global\let\ifmypprint\iffalse 
\def\mypprint{\global\let\ifmypprint\iftrue}
\global\let\iftorefs\iffalse
\def\torefs{\global\let\iftorefs\iftrue}
\global\let\dofloatfig\iffalse
\def\floatthefig{\let\dofloatfig\iftrue}
\mypprint 
\ifmypprint
\documentstyle[twocolumn,prl,aps]{revtex}
\floatthefig
\else
  \documentstyle[preprint,aps]{revtex}
  \iftorefs 
    \catcode`\@=11 
    
    \def\figure{\let\@capwidth\columnwidth\@float{figure}}
    \let\endfigure\end@float
    \@namedef{figure*}{\let\@capwidth\textwidth\@dblfloat{figure}}
    \@namedef{endfigure*}{\end@dblfloat}
    \catcode`\@=12 
     \floatthefig 
  \fi
\fi

\input epsf.tex
\begin{document}
\draft
\title{Onset of Surface-Tension-Driven B\'{e}nard Convection}
\author{Michael F. Schatz\cite{mfsemail}, Stephen J. VanHook, 
William D. McCormick, \\
J. B. Swift, and Harry L. Swinney\cite{hlsemail}}
\address{Center for Nonlinear Dynamics and Department of Physics\\
The University of Texas at Austin, Austin, Texas 78712}
\date{\today}
\maketitle
\widetext
\begin{abstract}
Experiments with shadowgraph visualization reveal a subcritical 
transition to a hexagonal
convection pattern in thin liquid layers that have a free upper
surface and are heated from below.  The measured critical Marangoni 
number (84) and observation of hysteresis ($3\%$) agree with theory.  
In some experiments, imperfect bifurcation 
is observed and is attributed to deterministic forcing caused in part by
the lateral boundaries in the experiment.  

\end{abstract}


\narrowtext

The onset of motion in heated fluid layers with a free 
upper surface has eluded complete understanding ever since 
B\'{e}nard's investigation
\cite{benard1900} of these flows established 
thermal convection as a paradigm for pattern formation in 
nonequilibrium systems \cite{CandH}. 
Rayleigh's analysis \cite {Rayleigh16} of this problem assumed that 
buoyancy effects, which are always present in layers heated from 
below, caused convection, but the threshold that Rayleigh predicted 
did not agree with B\'{e}nard's observations.  
Forty years elapsed before it was recognized 
that the instability observed 
in B\'{e}nard's studies was caused not by buoyancy but by surface tension 
gradients 
\cite {Block56}, as characterized by the Marangoni number $M$ 
(see Fig.\ \ref {sketch}).
Linear theory \cite{Pearson58} yields onset at $M_{c}$=$80$.  
Weakly nonlinear theory 
\cite{SandS,CandL} predicts a subcritical (hysteretic) transition to a 
hexagonal pattern.  Only 
a single experimental investigation \cite{KandB} has systematically 
examined the onset of convection for layers sufficiently thin such that 
surface tension forces dominate over buoyancy.  That experiment revealed a 
primary transition to a concentric roll pattern at values of $M$ that 
decreased as the fluid layers became thinner; for the thinnest layers 
studied, rolls emerged at $M$ an order of magnitude smaller 
than $M_{c}$ from theory.

	In this Letter we present evidence for a 
well-defined primary transition in surface-tension-driven 
B\'{e}nard (Marangoni) convection 
experiments designed so that surface tension forces dominate over 
buoyancy to a greater
extent than in previous 
investigations.  We observe a hysteretic bifurcation to a defect-free array 
of hexagonal cells; this bifurcation is modeled by an 
amplitude equation, which permits comparison to both linear and
weakly nonlinear stability theory.  We also observe hexagons to 
arise from an imperfect bifurcation where the
hysteresis disappears; this bifurcation 
is described qualitatively with the addition of a 
deterministic forcing term to the amplitude equation.  
In our experiments, surface tension effects are 40 times larger
than buoyancy effects, {\it i.e.}, $M/R$$\approx$40, where 
the Rayleigh number is defined as 
$R$$\equiv$${{g}\beta}\bigtriangleup{\it T}{d}^3$/$\nu\kappa$ with 
\ifmypprint\pagebreak \vspace*{1.15in} \noindent \fi
liquid expansion coefficient $\beta$ and 
gravitational acceleration $g$.  A necessary 
condition for the flow to be 
surface-tension-dominated is $M/R$$\geq$1 \cite{DandH}; previous experiments
attained $M/R$$<$11 \cite{KandB}.  

\dofloatfig
\begin{figure}
\epsfxsize=3.4 truein
\centerline{\epsffile{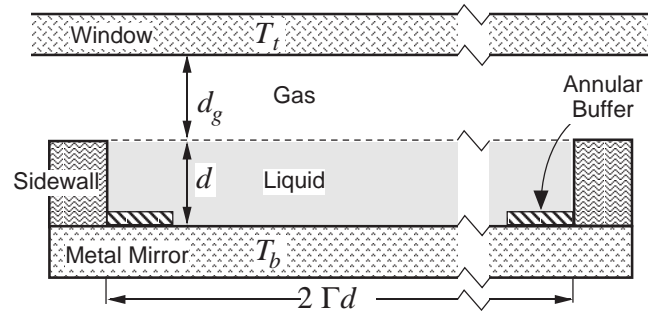}}
\smallskip
\caption[]{Cross section of our cylindrical convection cell.
The dimensionless control parameter is the
Marangoni number $M$$\equiv$$\sigma_{T}\bigtriangleup{T}{d}$/$\rho\nu\kappa$, 
where $\sigma_{T}$$\equiv$$\mid$${d}\sigma$/$dT$$\mid$, 
and $\sigma$, $\rho$, $\nu$, $\kappa$
are, respectively, the liquid surface tension, density, kinematic viscosity 
and thermal diffusivity.  The assumption of conductive heat 
transport is used
to obtain the mean temperature across the liquid layer 
$\bigtriangleup{\it T}$=$(T_{b}-T_{t})$/$(1+H^{-1})$ with the Biot number 
$H$=${\it k}_{g}{\it d}$/${\it k}{\it d}_{g}$ 
defined in terms of the thermal 
conductivities $k$ and ${\it k}_{g}$
of the liquid and gas, respectively.}

\label{sketch}
\end{figure} 	
\fi
	
	The experiments are performed on a purified silicone oil layer 
(${d}$=0.0419$\pm$0.0005 cm) that is bounded from below by 1-cm-thick 
gold-coated aluminum mirror (Fig.\ \ref{sketch}) \cite{McGregor}.  A 
uniform air layer
(${d}_{g}$=0.0455$\pm$0.0008 cm) lies between the oil layer and a 
1-mm-thick sapphire window.  The oil is confined by
a Teflon sidewall ring of inner diameter 4.53$\pm$0.01 cm.  
Because nonuniform wetting 
can cause large relative changes in the liquid layer thickness near the 
sidewall, a polyethersulfone annular buffer of thickness 50 $\mu$m suppresses
convection adjacent to the sidewall \cite{AMandC}.  The inner 
diameter of the buffer determines the 
radius to height ratio $\Gamma$=45.6$\pm$0.1 of the convecting region.  
The 
oil and the 
air layer depths each vary by less than 1$\%$ over the 
central 70$\%$ of 
the convecting region, as measured both mechanically and 
interferometrically.  Use of a purified \cite {HandS} silicone oil 
(96.7$\%$ hexacosamethyldodecasiloxane) avoids both 
condensation \cite {HandS} 
and cross-diffusive effects \cite{MandS} that can
affect pattern formation.  A temperature gradient is 
imposed by water-cooling 
the window to a temperature
${\it T}_{t}$=13.320$\pm$0.005$^\circ$C  
and by computer-controlled-heating 
the mirror to a temperature $T_{b}$ that
fluctuates less than $\pm$0.0005${^{\circ}}$C.  For    
sufficiently small ${\it T}_{b}-{\it T}_{t}$, the surface tension 
$\sigma$(${\it T}$) at the liquid-gas interface is uniform; 
however, with
${\it T}_{b}-{\it T}_{t}$ sufficiently large, instability causes surface 
tension variations that drive flow in the 
bulk.  The shadowgraph technique is used to 
detect onset and to visualize patterns.
Images are digitized and background subtracted to improve the 
signal-to-noise ratio.  The time scale in the experiment 
is set by the vertical diffusion time, $t_{v}$=${{\it d}}^{2}$/$\kappa$=1.9 s.

\dofloatfig
\begin{figure}
\epsfxsize=3.4 truein
\centerline{\epsffile{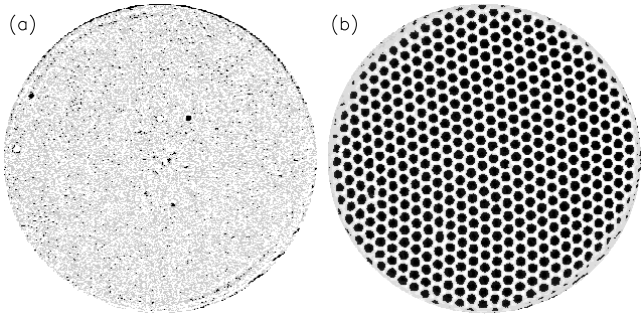}}
\medskip
\caption[]{The abrupt onset of hexagons in Marangoni convection. 
(a) Just prior to onset, weak convection rolls develop at the 
boundary for $\epsilon$=--5.6$\times$$10^{-3}$.  (b) A hexagonal
pattern fills the entire convection apparatus 
for $\epsilon$=--2.5$\times$$10^{-3}$.} 
\label{tohexes}
\end{figure} 
\fi
      
	Figure \ref{tohexes} demonstrates that the conductive state undergoes 
an abrupt transition to hexagons as $M$ is increased slowly 
(d${M}$/dt=$10^{-4}$ in units of ${t_{v}}^{-1}$).   
Just prior to onset, weak circular 
convection rolls arise near the boundary [upper left and lower right in 
Fig.\ \ref{tohexes}(a)]; we believe these rolls are driven by static 
forcing due to the slight mismatch of 
thermal conductivity between the annular buffer and the 
liquid.  Convection cells first 
appear within a portion of the boundary rolls after an increase
of $2\times10^{-3}$ in 
$\epsilon$$\equiv$$(M/M_{c})$--1.  ($M_{c}$ is determined
from the experiments, as will be described.)  Additional hexagons 
then nucleate from the initial cells
and propagate as a traveling front, invading the apparatus until the 
entire flow domain is filled with the hexagonal pattern
[Fig.\ \ref{tohexes}(b)].  
The resulting pattern is nearly free from defects since the 
lattice is grown from a single ``seed crystal'' at the boundary.  
The front propagates across the apparatus in 
approximately 900$t_{v}$, 
a time short compared to the horizontal diffusion time 
4$\Gamma^{2}t_{v}$$\approx$$8000t_{v}$.  

	On decreasing $\epsilon$ quasistatically, hexagons 
persist at parameter values below that for the first appearance 
of cells (Fig.\ \ref {tocond}).  The transition to the conductive state 
occurs gradually; the control parameter 
typically must be decremented through a range of $\epsilon$ before the 
hexagonal convection pattern 
disappears.
The front between hexagonal and conductive states can remain stationary
indefinitely (or at least for times long compared to the horizontal diffusion 
time) for states like Fig.\ \ref{tocond}(b) or (c) if $\epsilon$ is held 
constant. 

\dofloatfig
\begin{figure}
\epsfxsize=3.4 truein
\centerline{\epsffile{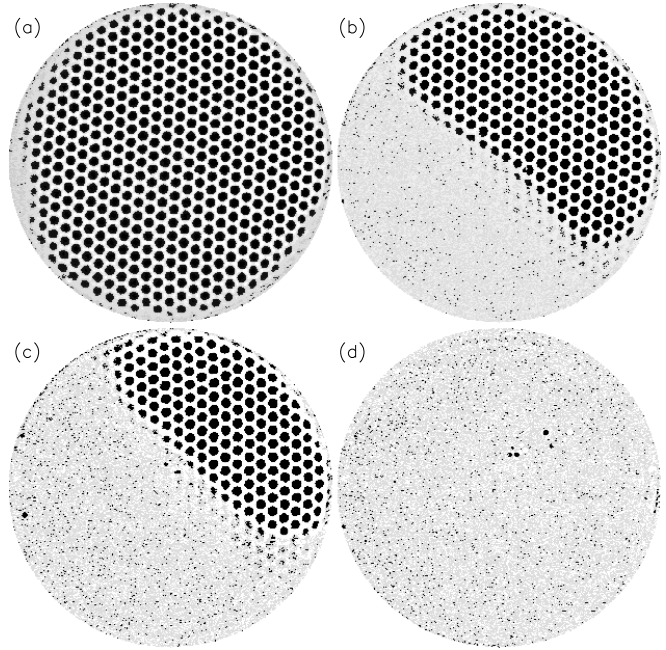}}
\smallskip
\caption{Return to the conductive state for decreasing $\epsilon$.  Below onset
hexagonal convection persists at (a) $\epsilon$=--2.60$\times$$10^{-2}$, 
(b) $\epsilon$=--2.82$\times$$10^{-2}$, and (c) 
$\epsilon$=--2.96$\times$$10^{-2}$ 
before disappearing at (d) $\epsilon$=--3.20$\times10^{-2}$.}
\label{tocond}
\end{figure} 
\fi

Near onset, the hexagonal pattern arises 
from the interaction of three plane wave (roll) solutions whose wavevectors
have a magnitude equal to the critical wavenumber and differ in angle 
by 2${\pi}$/3 \cite{Ciliberto}.  The evolution of the pattern can then 
be described by a 
Landau equation for the amplitude $A$ 

\begin{equation}
$$\dot{A} = {\epsilon}A + {\alpha}A^{2} - A^{3} + f,$$
\end{equation}

\noindent
with $\alpha$$>$0 and $f$ a constant that can account for 
deterministic forcing.  In some cases, 
the coefficients in Eq. (1) can be computed from the full fluid 
equations \cite {Davis87_Bestehorn}.   The existence of hexagons requires 
$\alpha$$\neq$0; thus, the bifurcation from the conductive state must be 
subcritical.  The 
solutions for hexagonal convection and for conduction are both linearly 
stable over a range of 
parameter:  $\epsilon_{\alpha}$$\leq$$\epsilon$$\leq{0}$ with 
$\epsilon_{\alpha}$=--$\alpha^{2}/4$ (the conductive state is linearly 
unstable for $\epsilon$$>$0).

Equation (1) is a variational model that exhibits relaxational time 
dependence governed by a potential 
function \cite {MNandT}, which we first consider for $f$=0.  
Over the parameter range where both 
conduction and convection are stable, each state corresponds to a 
minimum of the
potential; one state represents the global minimum while the other 
state, the metastable phase, represents a local minimum.  The potential
varies as $\epsilon$ changes; at a parameter value $\epsilon_{m}$ (the 
Maxwell point) both states have equal values of the potential. 
As $\epsilon$
passes through $\epsilon_{m}$, the states exchange the roles of global 
stability/metastability.  For Eq. (1), the conductive state is globally 
stable for 
$\epsilon$$<$$\epsilon_{m}$=$8\over{9}$$\epsilon_{\alpha}$ and metastable 
for $\epsilon_{m}$$<$$\epsilon$$<0$.

\dofloatfig
\begin{figure}
\epsfxsize=3.4 truein
\centerline{\epsffile{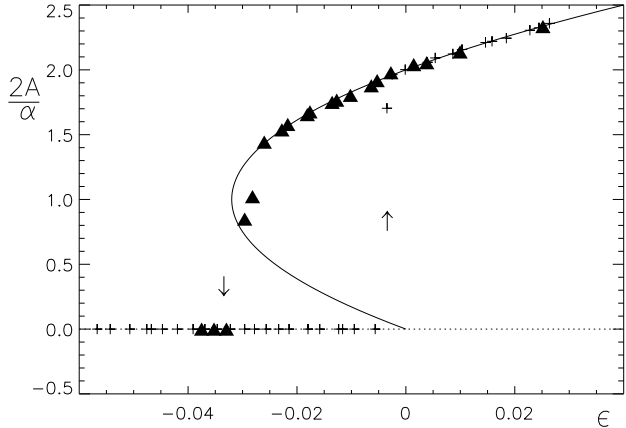}}
\smallskip
\caption[]{Hysteresis at the onset of Marangoni convection is demonstrated by
a plot of the Fourier mode amplitude $A$ from shadowgraph 
images $vs$ $\epsilon$.  Convection appears suddenly with 
slowly increasing $\epsilon$ ($+$) and persists 
below onset for slowly decreasing $\epsilon$ (triangles).  
A fit to the convective
branch (-----) yields $M_{c}$=83.6
($\bigtriangleup{T_{c}}$=1.65${^{\circ}}$C), which we use to compute 
$\epsilon$.}  
\label{amp}
\end{figure} 
\fi

To compare the experimental observations to the model, we compute 
two-dimensional spatial power spectra from shadowgraph images.  The spectra
are azimuthally averaged and normalized to the variance of the 
image intensity.  The mean position of the fundamental
spectral peak yields the wavenumber 1.90$\pm$0.02 
(nondimensionalized by $d$); linear stability analysis predicts a critical 
wavenumber of 
1.99 \cite {Pearson58}.  The wavenumber is independent
of $\epsilon$ 
for the range investigated.  The amplitude in Fig.\ \ref{amp} is the square 
root of the power contained in 
the spectral peak 
at the fundamental wavenumber.

	Figure \ref {amp} demonstrates that the experimental observations 
illustrated in Figs. \ref {tohexes} and 
\ref {tocond} are consistent with 
Eq. (1).  Hexagonal 
convection amplitudes for increasing and decreasing $\epsilon$ 
near the bifurcation are fit by a parabola, as suggested by (1) with 
$f$=0; from this fit we estimate $M_{c}$=$83.6$ with a precision of 
$\pm$0.5 in $M$.  The uncertainty in the accuracy is $\pm$11 in $M$, 
primarily due to 
the uncertainty in the thermal properties for the silicone oil.
From Fig.\ \ref{amp}, we also estimate 
$\epsilon_{\alpha}$=--3.2$\pm{0.3}$$\times10^{-2}$ 
and $\epsilon_{m}$=--2.8$\pm{0.3}$$\times10^{-2}$ \cite {Bodenschatz}.  For 
increasing $\epsilon$, 
the conductive state shown in Fig.\ \ref{tohexes}(a) is deep within the 
metastable regime when the initial onset
occurs.  The weak convection roll at the boundary 
provides a sufficient perturbation  
to push the 
system over the potential barrier, and the front between the two states 
propagates to spread the globally stable state (hexagons)
across the entire apparatus.  With decreasing $\epsilon$, hexagonal convection
can 
become metastable; however, the range of parameter values where hexagons are 
metastable is nearly 
an order of magnitude smaller than the region of metastability for conduction. 
This suggests that the transition back to conduction will be more sensitive 
to small 
spatial variations in $\epsilon$ due to nonuniformities in the depths of both 
liquid and gas layers.  Thus, for values of $\epsilon$ 
near the metastable region 
of hexagons, 
the front will move in stages to spread the
conductive state across the apparatus, as shown in Figs. 
\ref {tocond}(b) and \ref {tocond}(c). 

\dofloatfig
\begin{figure}
\epsfxsize=3.4 truein
\centerline{\epsffile{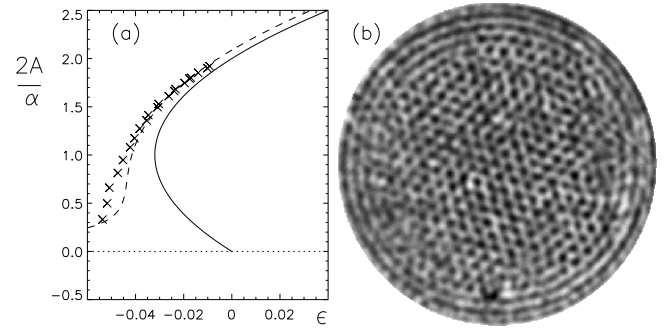}}
\smallskip
\caption[]{Imperfect bifurcation in Marangoni convection. (a)  Data ($\times$)
are compared to Eq. (1) with $f$=1.3$f_{c}$ (- - - -).  The 
parabola from Fig.\ 4 (-----) is also shown.  (b) 
Shadowgraph image of weak convective flow at 
$\epsilon$=--5.30$\times$$10^{-2}$  in the 
presence of significant deterministic forcing.}
\label{weakamp}
\end{figure} 
\fi

	In some cases, convection appears without 
hysteresis.  This situation arises, for example, in experiments where 
$\epsilon$ is repeatedly increased and decreased, causing conduction and 
convection to alternate.   As the number of cycles increases, hysteresis is 
observed at smaller values of $\epsilon$ and for a smaller range of $\epsilon$.
The convective onset occurs continuously [Fig.\ \ref{weakamp}(a)] after a 
sufficient number of cycles; this number varies from 3 to 15 between 
different experimental runs.  Our observations indicate that deterministic
forcing, which increases slowly (on a time scale much longer than 
the horizontal
diffusion time), causes an imperfect 
bifurcation \cite {Iooss90}.  Equation (1) models imperfect bifurcation 
with $f$$\not=$0; for $f$$>$$f_{c}$=$\alpha^{3}$/27, hysteresis 
disappears at 
the onset of hexagons.  In this regime, Eq. (1) qualitatively describes 
the amplitudes measured from our experiments [Fig.\ \ref{weakamp}(a)]; the 
difference between the model and the data suggests that a more complex form
for $f$ [{\it e.g.}, $f(\epsilon)]$ is necessary for 
quantitative agreement.

The physical origin of the forcing that causes imperfect bifurcation has 
not been determined definitively; however, the observation of low amplitude
rolls parallel to the lateral boundary [Fig.\ \ref{weakamp}(b)] 
suggests sidewall boundaries are affecting the flow.  
Similar structures arise
in buoyancy-driven convection with intentional thermal forcing at the 
sidewall \cite {AMandC}, although forcing for surface-tension-driven flow 
is probably more complex because some deformation of the free 
surface at the boundary 
is unavoidably present due to nonuniform contact 
line pinning at the sidewall.  The low
amplitude flows become increasingly cellular away from the boundaries
and toward the
center of the apparatus; moreover, with increasing $\epsilon$ above
onset, the rolls at the sidewall are supplanted by hexagonal cells as the 
amplitudes arising from imperfect bifurcation approach the amplitudes 
observed during hysteretic onset [Fig.\ \ref {weakamp}(a)].

Our experimental studies of onset confirm the predictions of theory and 
suggest an explanation
for the puzzling disagreement between previous experiments and theory for 
the century-old problem of surface-tension-driven B\'{e}nard 
convection.  Our determination of $M_{c}$=84 
is in reasonable agreement with $M_{c}$ from linear 
theory \cite {Nield64}, and our observation of subcritical bifurcation is in
accord with weakly nonlinear theory.  Our finding of 
3.2$\%$ hysteresis sets a standard for comparison to nonlinear theories, 
whose estimates of hysteresis range from 0.2$\%$ \cite {CandL} 
to 2.3$\%$ \cite {SandS}.  Observation of imperfect bifurcation 
demonstrates the sensitivity of the primary 
instability in Marangoni convection to perturbations; the appearance of rolls 
before hexagons at $M$$<<$$M_{c}$ in previous experiments \cite{KandB} may
well be due to this sensitivity.
 
	We are grateful to M. Bestehorn, S. H. Davis, R. E. Kelly, and 
E. L. Koschmieder for helpful discussions.  This work is supported by a 
NASA Microgravity Science and Applications Division Grant No. 
NAG3-1382.  One of us (S.J.V.H.) acknowledges support from 
the NASA Graduate Student 
Researchers Fellowship Program.


%
\dofloatfig\else
\begin{figure}
\smallskip
\caption[]{Cross section of our cylindrical convection cell.
The dimensionless control parameter is the
Marangoni number $M$$\equiv$$\sigma_{T}\bigtriangleup{T}{d}$/$\rho\nu\kappa$, 
where $\sigma_{T}$$\equiv$$\mid$${d}\sigma$/$dT$$\mid$, 
and $\sigma$, $\rho$, $\nu$, $\kappa$
are, respectively, the liquid surface tension, density, kinematic viscosity 
and thermal diffusivity.  The assumption of conductive heat 
transport is used
to obtain the mean temperature across the liquid layer 
$\bigtriangleup{\it T}$=$(T_{b}-T_{t})$/$(1+H^{-1})$ with the Biot number 
$H$=${\it k}_{g}{\it d}$/${\it k}{\it d}_{g}$ 
defined in terms of the thermal 
conductivities $k$ and ${\it k}_{g}$
of the liquid and gas, respectively.}
\label{sketch}
\end{figure} 	
\begin{figure}
\medskip
\caption[]{The abrupt onset of hexagons in Marangoni convection. 
(a) Just prior to onset, weak convection rolls develop at the 
boundary for $\epsilon$=--5.6$\times$$10^{-3}$.  (b) A hexagonal
pattern fills the entire convection apparatus 
for $\epsilon$=--2.5$\times$$10^{-3}$.} 
\label{tohexes}
\end{figure} 
\begin{figure}
\smallskip
\caption{Return to the conductive state for decreasing $\epsilon$.  Below onset
hexagonal convection persists at (a) $\epsilon$=--2.60$\times$$10^{-2}$, 
(b) $\epsilon$=--2.82$\times$$10^{-2}$, and (c) 
$\epsilon$=--2.96$\times$$10^{-2}$ 
before disappearing at (d) $\epsilon$=--3.20$\times10^{-2}$.}
\label{tocond}
\end{figure} 
\begin{figure}
\smallskip
\caption[]{Hysteresis at the onset of Marangoni convection is demonstrated by
a plot of the Fourier mode amplitude $A$ from shadowgraph 
images $vs$ $\epsilon$.  Convection appears suddenly with 
slowly increasing $\epsilon$ ($+$) and persists 
below onset for slowly decreasing $\epsilon$ (triangles).  
A fit to the convective
branch (-----) yields $M_{c}$=83.6
($\bigtriangleup{T_{c}}$=1.65${^{\circ}}$C), which we use to compute 
$\epsilon$.}  
\label{amp}
\end{figure} 
\begin{figure}
\smallskip
\caption[]{Imperfect bifurcation in Marangoni convection. (a)  Data ($\times$)
are compared to Eq. (1) with $f$=1.3$f_{c}$ (- - - -).  The 
parabola from Fig.\ 4 (-----) is also shown.  (b) 
Shadowgraph image of weak convective flow at 
$\epsilon$=--5.30$\times$$10^{-2}$  in the 
presence of significant deterministic forcing.}
\label{weakamp}
\end{figure} 
\fi
\ifmypprint\else
 \vfill\eject
 \epsfxsize=6 truein  \centerline{\epsffile{prl_benard_fig1.ps}}
  \vfill{\small Schatz {\it et al.}, Fig.~1} \vfill\eject
 \epsfysize=6.0truein  \centerline{\epsffile{prl_benard_fig2.ps}}
 \vfill{\small Schatz {\it et al.}, Fig.~2} \vfill\eject \epsfysize=6truein
 \centerline{\epsffile{prl_benard_fig3.ps}}
 \vfill{\small Schatz {\it et al.}, Fig.~3} \vfill\eject
 \epsfxsize=6truein \centerline{\epsffile{prl_benard_fig4.ps}}
 \vfill{\small Schatz {\it et al.}, Fig.~4} \vfill\eject
 \epsfxsize=6truein \centerline{\epsffile{prl_benard_fig5.ps}}
 \vfill{\small Schatz {\it et al.}, Fig.~5} \vfill\eject
\fi

%
%


\begin{references}
\bibitem[*]{mfsemail}
Electronic mail:  schatz@chaos.ph.utexas.edu
\bibitem[\dag]{hlsemail}
Electronic mail:  swinney@chaos.ph.utexas.edu
\bibitem{benard1900}
H. B\'{e}nard, Rev. G\'{e}n. Sci. Pure Appl. {\bf 11}, 1261 (1900).
\bibitem{CandH}
M. C. Cross and P. C. Hohenberg, Rev. Mod. Phys. {\bf 65}, 851 (1993).
\bibitem{Rayleigh16}
Lord Rayleigh, Phil. Mag. {\bf 32}, 529 (1916).
\bibitem{Block56}
M. J. Block, Nature {\bf 178}, 650 (1956).
\bibitem{Pearson58}
J. R. A. Pearson, J. Fluid Mech. {\bf 4}, 489 (1958).
\bibitem{SandS}
J. Scanlon and L. Segal, J. Fluid Mech. {\bf 30}, 149 (1967).
\bibitem{CandL}
A. Cloot and G. Lebon, J. Fluid Mech. {\bf 145}, 447 (1984).
\bibitem{KandB}
E. L Koschmieder and M. I. Biggerstaff, J. Fluid Mech. {\bf 167}, 49 (1986).
\bibitem{DandH}
S. H. Davis and G. M. Homsy, J. Fluid Mech. {\bf 98}, 527 (1980).  
\bibitem{McGregor}
Physical 
properties for the liquid and the gas ($\rho$=$0.925$ g/cm$^{3}$, 
$\nu$=$7.1$ cS, 
$\sigma_{T}$=$0.068$ dyn/cm${^{\circ}}$C,
$k$=1.25$\times$$10^{4}$ erg/cm$\cdot$s${^{\circ}}$C,
$\beta$=1.1$\times$$10^{-3}$$^{\circ}$C$^{-1}$, 
$\kappa$=8.6$\times{10^{-4}}$ cm$^2$/s, and 
$k_{g}$=2.55$\times$$10^{3}$ erg/cm$\cdot$s${^{\circ}}$C) are obtained 
from the literature; see, for example, R. R. McGregor, 
{\it Silicones and Their Uses}
(McGraw-Hill, New York, 1956) and references therein; {\it CRC Handbook of 
Chemistry and Physics, 59th ed.} (CRC Press, West Palm Beach, 1978); H. J.
Palmer and J. C. Berg, J. Fluid Mech. {\bf 47}, 779 (1971).
\bibitem{AMandC}
G. Ahlers, C. W. Meyer and D. S. Cannell, J. Stat. Phys. {\bf 54}, 1121 (1989).
\bibitem{HandS}
M. F. Schatz and K. Howden, submitted to Exp. Fluids (1995).
\bibitem{MandS}
E. Moses and V. Steinberg, Phys. Rev. Lett. {\bf 57}, 2018 (1986).
\bibitem{Ciliberto} 
S. Ciliberto, P. Coullet, J. Lega, E. Pampaloni, and C. P\'{e}rez-Garc\'{i}a,
 Phys. Rev. Lett. {\bf 65}, 2370 (1990).
\bibitem{Davis87_Bestehorn}
S. H. Davis, Annu. Rev. Fluid Mech. {\bf 19}, 403 (1987); M. Bestehorn, 
Phys. Rev. E {\bf 48}, 3622 (1993).
\bibitem{MNandT}
B. A. Malomed, A. A. Nepomnyashchy, and M. I. Tribelshy, Phys. Rev. 
A {\bf 42}, 7244 (1990).
\bibitem{Bodenschatz}
$\epsilon_{\alpha}$ is typically an 
order of magnitude smaller for the onset of hexagons in 
buoyancy-driven convection; see, for example, 
E. Bodenschatz, J. R. de Bruyn, G. Ahlers and D. S. Cannell, 
Phys. Rev. Lett. {\bf 67}, 3078 (1991).
\bibitem{Iooss90}
G. Iooss and D. D. Joseph, {\it Elementary Stability and Bifurcation Theory} (Springer-Verlag, Berlin, 1990).
\bibitem{Nield64}
Linear theory that accounts for the small effects of both buoyancy 
($M/R$=39) and heat transfer ($H$=0.2--see Fig. 1) predicts $M_{c}$=86.9; see 
D. A. Nield, J. Fluid Mech. {\bf 19}, 341 (1964).
\end{references}
\end{document}